\documentclass[showpacs,twocolumn,pra,superscriptaddress,notitlepage]{revtex4-2}
\usepackage{qcircuit}
\usepackage[dvips]{graphicx}
\usepackage{amsmath,amssymb,amsthm,mathrsfs,amsfonts,dsfont}
\usepackage{subfigure, epsfig}
\usepackage{braket}
\usepackage{bm}
\usepackage{fancyhdr}
\usepackage{enumerate}
\usepackage{color}
\usepackage[dvipsnames]{xcolor}

\usepackage{hyperref}
\hypersetup{colorlinks=true, linkcolor=blue, citecolor=blue, urlcolor=black }

\newcommand{\comments}[1]{}

\usepackage{graphicx}
\usepackage{amsmath}
\usepackage{amsfonts}
\usepackage{gensymb}
\usepackage{tabularx}
\usepackage[linesnumbered, ruled, vlined]{algorithm2e}

\SetKwRepeat{Do}{do}{while}

\newcommand{\HM}{$\rm{H}_{2}$}
\newcommand{\Alkene}{$\rm{C}_{6}\rm{H}_{8}$ }
\newcommand{\HChain}{$\rm{H}_{10}$}
\newcommand{\CRing}{$\rm{C}_{18}$ }
\newcommand{\TheMethod}{DMET-ESVQE}
\newcommand{\mol}[1]{\mathrm{#1}}

\def\ket#1{\mathinner{|{#1}\rangle}}
\def\braket#1{\mathinner{\langle{#1}\rangle}}

\newcommand{\bytedance}{ByteDance Inc, Zhonghang Plaza, No. 43, North 3rd Ring West Road, Haidian District, Beijing.}
\newcommand{\tsinghua}{Department of Chemistry, Tsinghua University,  Beijing 100084, China}
\newcommand{\cas}{Institute of Computing Technology, Chinese Academy of Sciences}
\newcommand{\ucas}{University of Chinese Academy of Sciences}
\newcommand{\peking}{Center on Frontiers of Computing Studies, Peking University, Beijing 100871, China}

\newcommand{\oxford}{Clarendon Laboratory, University of Oxford, Oxford OX1 3PU, UK}

\begin{document}

\title{Toward {Practical} Quantum Embedding Simulation of Realistic Chemical Systems on Near-term Quantum Computers}

\date{\today}

\author{Weitang Li}
\affiliation{\bytedance}
\affiliation{\tsinghua}

\author{\mbox{Zigeng Huang}}
\affiliation{\bytedance}

\author{Changsu Cao}
\affiliation{\bytedance}

\author{Yifei Huang}
\affiliation{\bytedance}

\author{\mbox{Zhigang Shuai}}
\affiliation{\tsinghua}

\author{Xiaoming Sun}
\affiliation{\cas}
\affiliation{\ucas}

\author{Jinzhao Sun}
\affiliation{\oxford}

\author{Xiao Yuan}
\affiliation{\peking}

\author{Dingshun Lv} \email{lvdingshun@bytedance.com}
\affiliation{\bytedance}

\begin{abstract}
Quantum computing has recently exhibited great potentials in predicting chemical properties for various applications in drug discovery, material design, and catalyst optimization. Progress has been made in simulating small molecules, such as LiH and hydrogen chains of up to 12 qubits, by using quantum algorithms such as variational quantum eigensolver (VQE). {Yet, originating from limitations of the size and the fidelity of near-term quantum hardware, how to accurately simulate large realistic molecules remains a challenge. 
Here, integrating an adaptive energy sorting strategy and a classical computational method, the density matrix embedding theory, which effectively finds a shallower quantum circuit and reduces the problem size, respectively, we show a means to circumvent the limitations and demonstrate the potential toward solving real chemical problems. We numerically test the method for the hydrogenation reaction of \Alkene and the equilibrium geometry of the \CRing molecule, with basis sets up to cc-pVDZ (at most 144 qubits). The simulation results show accuracies comparable to those of advanced quantum chemistry methods such as coupled-cluster or even full configuration interaction, while the number of qubits required is reduced by an order of magnitude (from 144 qubits to 16 qubits for the \CRing molecule) compared to conventional VQE.} Our work implies the possibility of solving industrial chemical problems on near-term quantum devices.

\end{abstract}

\maketitle

\section{Introduction}
Various methods based on the wave function theory, from the primary mean-field Hartree-Fock to high accuracy coupled-cluster singles-and-doubles and full configuration interaction methods, have been developed to simulate the many-electron molecular systems~\cite{WFT1, WFT2}. However,  owing to the exponential wall~\cite{Kohn99}, the exact treatment of those systems with more than hundreds of orbitals remains intractable for classical computers, hindering further investigations on large realistic chemical systems.
Quantum computing is believed to be a promising approach to overcome the exponential wall in quantum chemistry simulation~\cite{mcardle2018quantum, aspuru2005simulated, Garnet20}, which may potentially boost relevant fields such as material design and drug discovery.
Despite the great potential, fault-tolerant simulation of realistic molecules is still far beyond the current reach~\cite{reiher2017elucidating, li2019electronic, berry2019qubitization, von2021quantum, lee2021even}. 
In the present noisy intermediate-scale quantum~(NISQ) era~\cite{preskill2018quantum}, the variational quantum eigensolver~(VQE), as one of the most popular quantum-classical algorithms~\cite{o2016scalable, peruzzo2014variational, kandala2017hardware,arute2020hartree,cerezo2020variational,bharti2021noisy,mcardle2018quantum,huggins2021efficient,endo2020variational,endo2020hybrid,hempel2018quantum,yuan2020quantum,fujii2020deep,mcardle2018quantum,XU2021,nam2020ground,cao2021larger,fujii2020deep}, has been exploited to experimentally study molecules from $\mol{H_2}$ (2 qubits)~\cite{peruzzo2014variational}, $\mol{BeH_2}$ (6 qubits)~\cite{kandala2017hardware}, $\mol{H_2O}$ (8 qubits)~\cite{nam2020ground}, to $\mol{H_{12}}$ (12 qubits)~\cite{arute2020hartree}. Meanwhile, the largest scale numerical simulation is $\mol{C_2H_4}$ (28 qubits)~\cite{cao2021larger}. 

{However, realistic chemical systems with an appropriate basis generally involve hundreds or thousands of qubits, whether VQE with NISQ hardware is capable of solving any practically meaningful chemistry problem remains open. The main challenge owes to limitations on the size (the number of qubits) and the fidelity (the simulation accuracy) of NISQ hardware~\cite{bharti2021noisy,preskill2018quantum,cerezo2020variational,endo2020hybrid}. Specifically, it is yet hard to scale up the hardware size while maintaining or even increasing the gate fidelity. Experimentally, when we directly implement VQE on more than hundreds of qubits, the number of gates needed might become too large so that errors would accumulate drastically and error mitigation would require too many measurements to reach the desired chemical accuracy. 
}

Adaptive and hybrid classical-quantum computational methods provide more economical ways to potentially bypass the conundrum. 
On the one hand, adaptive VQE algorithms~\cite{Grimsley19,tang2021qubit,zhang2020lowdepth} can greatly reduce the circuit depth hence alleviate the limitation on the gate fidelity. 
On the other hand, noticing the fact that most quantum many-body systems have mixed strong and weak correlation, we only need to solve the strongly correlated degrees of freedom using quantum computing and calculate the remaining part at a mean-field level using classical computational method. Along this line, several {hybrid} methods have been proposed by exploiting different classical methods~\cite{maier2005quantum,sun2016quantum,RMPDMFT}, such as density matrix embedding theory~\cite{DMET2012,DMET2013,Wouters16,rubin2016hybrid}, dynamical mean field theory~\cite{Troyer16,rungger2020dynamical,Chen_2021}, tensor network~\cite{yuan2020quantum,orus2019tensor}, and perturbation theory~\cite{sun2021perturbative}.
Density matrix embedding is one of the representative embedding methods that have been theoretically and experimentally developed in several works~\cite{DMET2012,DMET2013,Wouters16,wu2019projected,rubin2016hybrid,kawashima2021efficient,tilly2021reduced,wu2020enhancing,cui2020,fertitta2019energy,wen2020,FT2020,Garnet20}, yet the practical realization toward realistic chemical systems remains a significant technical challenge.

In this work, we integrate the adaptive energy sorting strategy~\cite{fan2021circuit} and density matrix embedding theroy~\cite{DMET2012,DMET2013,Wouters16,rubin2016hybrid,tilly2021reduced}, and provide a systematic way with multiscale descriptions of quantum systems toward practical quantum simulation of realistic molecules. 
We numerically study chemical systems with strong electron-electron correlation at specific geometries, including the homogeneous stretching of \HChain \ chain, the reaction energy profile for the hydrogenation of \Alkene with \HM~and the potential energy curve of the \CRing molecule~\cite{Anderson19}. While our method only uses a smaller number qubits (from 144 qubits to 16 qubits for the \CRing molecule) with a shallower quantum circuit, it reaches high accuracy which is comparable to coupled cluster or even full configuration interaction calculations. Our work reveals the possibility of studying realistic chemical processes on near-term quantum devices.

\section{Framework}

The generic Hamiltonian of a  quantum chemical system under Born-Oppenheimer approximation~\cite{mcardle2018quantum}  in second-quantized form can be expressed as
\begin{equation}
    \label{eq-ham}
    \begin{aligned}
    \hat{H}& = E_{\textrm{nuc}} + \sum_{k,l} \hat{D}_{kl} + \sum_{k,l,m,n}\hat V_{klmn},
    \end{aligned}
\end{equation}
where $E_{\textrm{nuc}}$ is the scalar nuclear repulsion energy, $\hat{D}_{kl}=d_{kl} \hat{a}^{\dagger}_{k} \hat{a}_{l}$ and $\hat{V}_{klmn}={\frac{1}{2} h_{klmn} \hat{a}^{\dagger}_{k} \hat{a}^{\dagger}_{l} \hat{a}_{m} \hat{a}_{n}}$ are the one and two-body interaction operator, respectively, 
$\hat{a}_{p}$ ($\hat{a}^{\dagger}_{p}$) is the fermionic  annihilation (creation) operator to the $p$th orbital, and $\{d_{kl}\}$ and $\{h_{klmn}\}$ are the corresponding one- and two-electron integrals calculated by classical computers, respectively. Here, we denote the spin-orbitals of the molecule by $k,l,m,n$.
To find a ground state of the Hamiltonian in Eq.~\eqref{eq-ham}, variational quantum eigensolvers (VQE) can be used in this task~\cite{cerezo2020variational,mcardle2018quantum}.
The key idea is that the parametrized quantum state $\Psi(\vec{\theta})$ is prepared and measured on a quantum computer, while the parameters are updated by a classical optimizer in a classical computer.
The ground state can be found by minimizing the total energy with respect to the variational parameters $\vec{\theta}$, following the variational principle,
$
	E = \min_{\vec{\theta}} {\langle\Psi(\vec{\theta})|\hat{H}|\Psi(\vec{\theta})\rangle}
$.

The above quantum algorithm entails the number of qubits no smaller than the system size, making it inaccessible to the  large realistic molecular systems.
Here, we introduce the quantum embedding approach to reduce the required quantum resources, originally proposed in Ref.~\cite{DMET2012}.
We consider to divide the total Hilbert space $\mathcal{H}$ of the quantum system into two subsystems, the fragment $A$ with ${L_A}$ bases $\{\ket{A_i}\}$ and environment $B$ with ${L_B}$ bases $\{\ket{B_j}\}$, respectively. 
The full quantum state in the bases of subsystems can be represented by $\ket{\Psi} = \sum_{i,j} \Psi_{ij} \ket{A_i}\ket{B_j}$ in a Hilbert space of dimension ${L_A \times L_B}$. However, this can be largely reduced by considering the entanglement between two subsystems. More specifically, the quantum state $\ket{\Psi}$ can be decomposed into the rotated basis of subsystems as $\ket{\Psi}  = \sum_{\alpha}^{L_A} \lambda_{\alpha} \ket{\tilde A_{\alpha}}\ket{\tilde B_{\alpha}}$, where the states $|\tilde{B}_{\alpha}\rangle$ can be regarded as the bath states. 
After the decomposition we indeed split the environment into at most $L_A$ bath states that are entangled with the fragment and the purely disentangled ones.
We could thus construct the embedding Hamiltonian by projecting the full Hamiltonian $\hat H \subset \mathcal{H}$ into the space spanned by the basis of fragment and bath as $\hat H_{\rm{emb}} = \hat P \hat H \hat P$ with the projector $\hat  P$ defined as $
\hat P=\sum_{\alpha \beta}|\tilde{A}_{\alpha} \tilde{B}_{\beta}\rangle\langle\tilde{A}_{\alpha} \tilde{B}_{\beta}|.
$
We note that the embedding Hamiltonian can be represented in the rotated spin-orbitals $p,q,r,s$ with renormalized coefficients $\tilde d_{p,q}$ and $\tilde h_{p,q,r,s}$ (See Appendix~\ref{appendix:DMET-construction}), and admits the second-quantized form as that in Eq.~\eqref{eq-ham}.

We can find that if $|\Psi\rangle$ is the ground state of a Hamiltonian $\hat H$,  it must also be the ground state of $\hat H_{\rm{emb}}$. This indicates that the solution of a small embedded system is the exact equivalent to that of the full system~\cite{Wouters16}, with the dimension of the embedded system reduced to ${L_A\times L_A }$.
In principle, the construction of $\hat P$ requires the exact ground state of the full system $\ket{\Psi}$, which makes it unrealistic from theory.
However, since we are interested in the ground state property, (for instance, the energy, which is a local density), we can consider to match the density or density matrix of the embedding Hamiltonian and the full Hamiltonian at a self-consistency level.
More specifically, we consider a set of coupled eigenvalue equations 
\begin{equation}
\hat H_{\rm{mf}} \ket{\Phi} = E_{\rm{mf}} \ket{\Phi},~ \hat H_{\rm{emb}} \ket{\Psi} = E_{\rm{emb}} \ket{\Psi},
\label{eq:eig_value_eq}
\end{equation}
which describe a low-level  mean-field system and a high-level interacting embedding system, respectively. 
Here, the mean-field Hamiltonian can be constructed provided the correlation potential $\hat{C}$ as $\hat H_{\rm{mf}} = \sum_{kl}\hat{D}_{kl}+\hat{C}$, and we can efficiently obtain the low-level wavefunction $\ket{\Phi}$ and hence the one-body reduced density matrix $^1D_{pq}=\braket{\hat{a}_p^{\dag}\hat{a}_q}$. 
Note that the embedding Hamiltonian can be constructed by the bath states, which is determined from  $\ket{\Phi}$. Therefore, the correlation potential $\hat{C}$ enters into the interacting theory from the projection $\hat P$.
Given the solution of the eigenvalue equations in Eq.~\eqref{eq:eig_value_eq},  we can also obtain the reduced density matrix of the embedded system. At a self-consistency level, we can match the reduced density matrices of the multilevel systems by adjusting the correlation potential $\hat{C}$ in the mean-field Hamiltonian $\hat H_{\rm{mf}}$, and we obtain a guess for the ground state solution at convergence. 

Next, we discuss how to get the solution of the high-level embedding Hamiltonian using variational quantum quantum eigensolvers.
The key ingredient in VQE is to design a proper circuit ansatz to approximate the unknown ground state of the chemical system. Here, we use the unitary coupled-cluster (UCC) ansatz~\cite{ucc-1, ucc-2, ucc-3}, which effectively considers the excitations and de-excitations above a reference state.  The UCC ansatz is defined as 
$
    |\Psi\rangle = \exp{(\hat{T} - \hat{T}^{\dagger})}|\Psi_{0}\rangle,
$
where $|\Psi_{0}\rangle$ is chosen as Hartree-Fock ground state {represented in the basis of} the embedded system, and $\hat{T}$ is the cluster operator.
The cluster operator truncated at single- and double-excitations has the form
\begin{equation}
\label{eq-excitation-operators}
    \hat T(\vec{\theta}) = \sum_{\substack{p\in vir\\r\in occ}} {\theta_{pr} \hat{T}_{pr}} + \sum_{\substack{p>q,r>s:\\p,q\in vir\\r,s\in occ}} {\theta_{pqrs} \hat{T}_{pqrs}}, \nonumber
\end{equation}
where the one- and two-body terms are defined as
$
    \hat{T}_{pr} = \hat{a}^{\dagger}_{p} \hat{a}_{r}
$ and
$
    \hat{T}_{pqrs} = \hat{a}^{\dagger}_{p} \hat{a}^{\dagger}_{q} \hat{a}_{r} \hat{a}_{s}
$, respectively.
Then, we can get the high-level wavefunction by optimizing the energy of the embedded system, $E = \min_{\theta} \braket{\Psi(\vec \theta)|\hat H_{\textrm{emb}}| \Psi(\vec \theta)}$, and thus can obtain the reduced density matrices ${^1 D}$. By matching the reduced density matrices with those of the mean-field system, this forms a self-consistency loop until convergence.

\section{Implementation}
\begin{figure}
\centering
  \includegraphics[width=\linewidth]{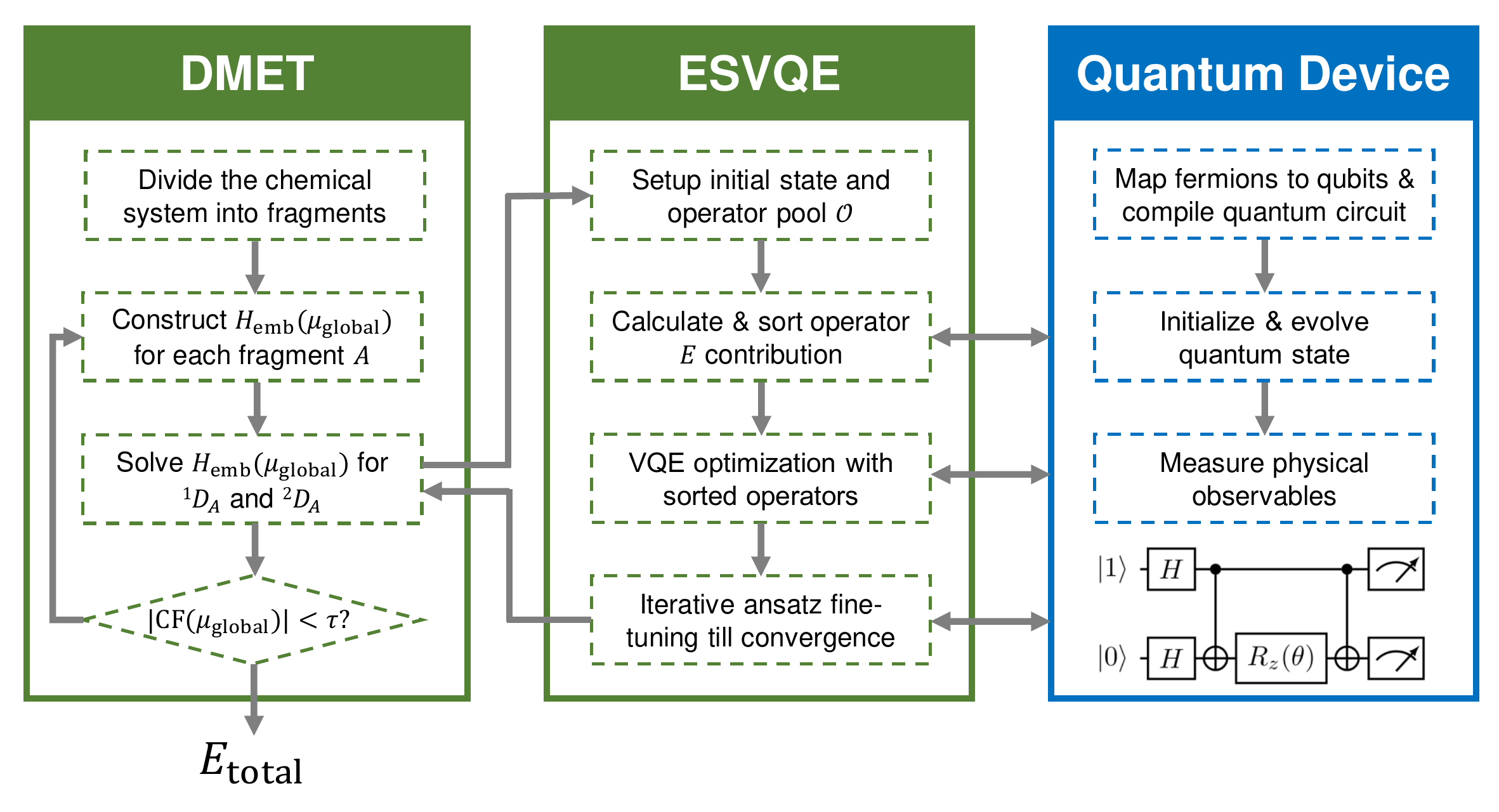}
  \caption{The workflow for the \TheMethod~method. The chemical system
  is first decomposed into fragments. Then the effective embedding Hamiltonian
  $H_{\rm{emb}}(\mu_{\textrm{global}})$ in DMET iteration is solved by ESVQE. 
  The ESVQE module 
  utilizes quantum devices in the blue box to prepare quantum states and measure physical observables. Both DMET iteration and ESVQE parameter optimization 
  are carried out on a classical computer, indicated by green boxes.}
  \label{fig:flowchart}
\end{figure}

Here, we discuss the implementation of the quantum embedding theory in practice.
In this work, we employ the energy sorting strategy~(ES) proposed in~Ref.~\cite{fan2021circuit} to select only the dominant excitations in the original operator pool and construct a compact quantum circuit in VQE procedures. 
An overall schematic flowchart for DMET-ESVQE, including both the DMET algorithm and the ESVQE solver, is presented in Fig.~\ref{fig:flowchart}. 
A complex chemical system is first decomposed into fragments by DMET {in a bootstrap manner} and each fragment is solved by ESVQE to obtain the reduced density matrices.
The DMET iteration is carried out on classical computers indicated by the green box.

In practical implementation for molecular systems, instead of partitioning the fragment in terms of the atoms, we determine the fragment partition based on the basis of atomic orbitals, such that coupling of the atomic orbitals could be captured in a more natural way. 
Once the partition is determined, the set of single-particle basis for fragment $A$, $\Omega^A=\{\ket{\phi^A_i}\}$, is often chosen as $\Omega^A = \bigcup_{j} \Omega_{j}^A$, where $\Omega_{j}^A$ is the set of basis located on the $j$th atom of fragment $A$.
However, such straight-forward scheme is inefficient for NISQ devices especially when large basis set is incorporated.
Here, we use a reduced basis set $\Omega^A = \bigcup_{j} \Omega_{j}^A \setminus  \Omega^A_{\textrm{mf}}$,
where $\Omega^A_{\textrm{mf}}$ is a set of inactive orbitals treated at mean-field level and 
excluded from the DMET iteration, and $\Omega^A_{\textrm{mf}}$ could be an empty set.
The resulting $\Omega^A$ can effectively capture the  entanglement between the orbitals and is more compact for the VQE procedure.
During the DMET optimization, we introduce a global chemical potential $\mu_{\rm global}$ to preserve the total number of electrons $N_{\rm occ}$, 
and the DMET cost function $\mathcal{L}(\mu_{\textrm{global}})$
can now be written as
\begin{equation}
\label{eq-cf-mu}
\begin{aligned}
    \mathcal{L} &\left(\mu_{\rm global}\right) = \\ 
    &\left(\sum_{A} \sum^{L_A}_{r \in \Omega_A} {^{1}D}_{r r}^{{\rm frag},A}\left(\mu_{\rm global}\right) + N_{\textrm{mf}} - N_{\rm occ}\right)^2,
\end{aligned}
\end{equation}
where
$N_{\textrm{mf}} = \sum_A \sum_{r \in \Omega^A_{\textrm{mf}} }{^{1}D}_{r r}^{\rm{mf}}$
is the number of electrons in the inactive orbitals
obtained at the mean-field level, termed as the single-shot embedding~\cite{Wouters16}.
We note that ${^{1}D}_{r r}^{\rm{mf}}$ is irrelevant to the self-consistency condition for DMET,
and thus $N_{\textrm{mf}}$ is not a function of $\mu_{\textrm{global}}$.
This feature distinguishes the the approach from simply adopting an active space high-level solver. More details can be found in Appendix~\ref{appnedix:DMET-constraint}.

For the ESVQE part, an efficient ansatz for each of the fragment is constructed by selecting dominant excitation operators in the operator pool $\mathcal{O}=\{    \hat{T}_{pr}, \hat{T}_{pqrs}\}$.
Here, the importance of the operator $\hat{T}_{i} \in \mathcal{O}$ is evaluated by the energy difference between the reference state as $\Delta E_{i} = E_{i}-E_{\textrm{ref}}$ with $E_i = \min_{\theta_i} \braket{\Psi_{\textrm{ref}}|e^{-\theta_i (\hat T_i - \hat T_i^\dagger)}\hat{H}e^{\theta_i (\hat T_i - \hat T_i^\dagger)}|\Psi_{\textrm{ref}}} $ and $E_{\textrm{ref}} =   \braket{\Psi_{\textrm{ref}}| \hat{H} |\Psi_{\textrm{ref}}}$. The operators with contributions above a threshold $|\Delta E_{i}| > \varepsilon$ are picked out and used to perform the VQE optimization. 
Extra fine-tuning can be performed by iteratively adding more operators to the ansatz until the energy difference $E^{(k-1)}-E^{(k)}$ between the ($k-1$)th and the $k$th iteration is smaller than a certain convergence criterion. In this work we skip this step for simplicity.
The overall procedure effectively reduces the resource requirements for the quantum devices.

In the DMET-ESVQE scheme, the number of qubits for the quantum solver is determined by the maximum number of orbitals in each of the fragment. 
Suppose in fragment $A$ there are $L_A$ spin orbitals and accordingly there are $L_A$ spin orbitals for the bath,
then under the fermion-to-qubit mapping (such as Jordan-Wigner transformation) $2L_A$ qubits are required for the quantum solver.

\section{Results and Discussion}
\label{sec:res}
\begin{figure}
\centering
  \includegraphics[width=\linewidth]{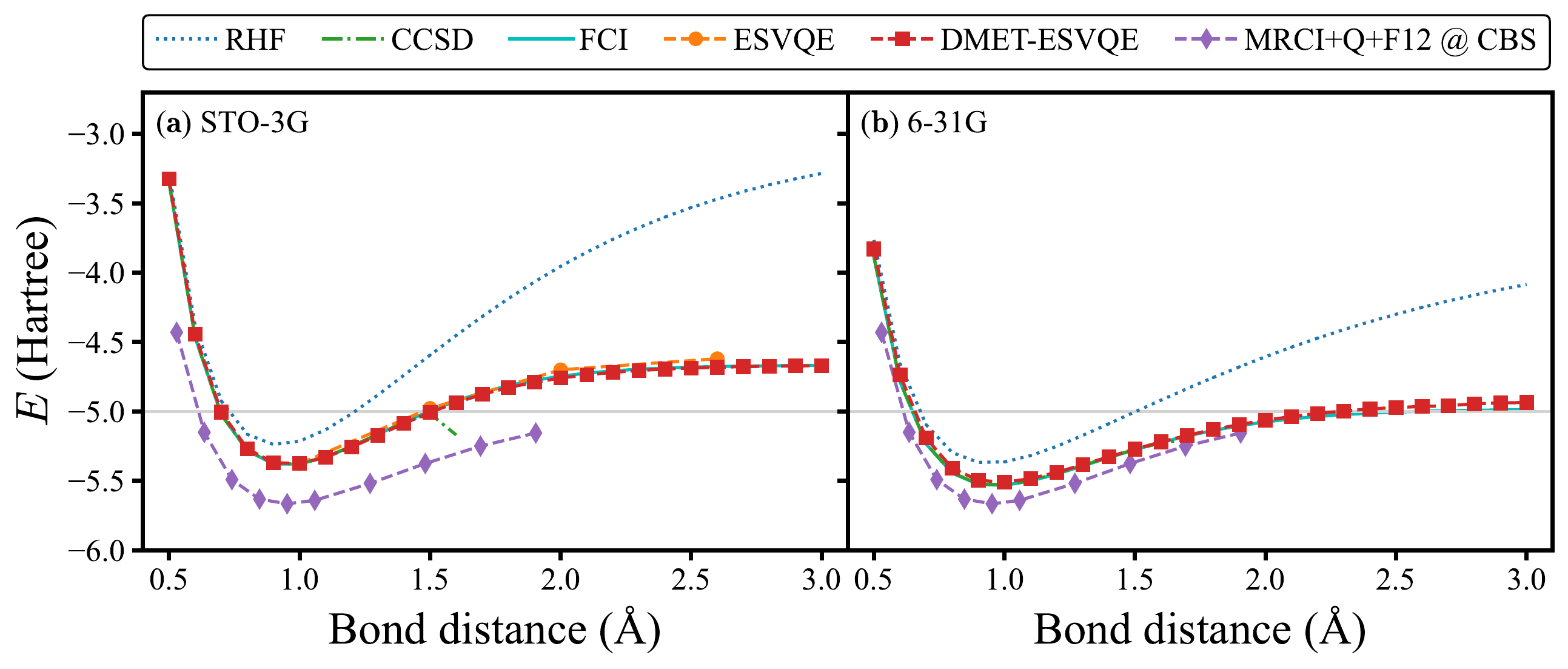}
  \caption{\TheMethod~simulated homogeneous stretching of a evenly-spaced hydrogen chain composed of 10 atoms in (a) STO-3G and (b) 6-31G basis set, in comparison with RHF, CCSD and FCI results.
  The MRCI+Q+F12@CBS results in both panels can be considered as the exact reference in the complete basis set (CBS) limit~\cite{White17}.
  For the STO-3G basis set, we also show the results obtained by conventional ESVQE.
  The grey horizontal line indicates the exact dissociation limit composed of non-interacting hydrogen atoms.
  }
  \label{fig:H10}
\end{figure}

To benchmark our algorithm, we first show the simulated potential energy curve for the homogeneous stretching of a hydrogen chain composed of 10 atoms in Fig.~\ref{fig:H10}, which is widely used as a benchmark platform for advanced many-body computation methods~\cite{Garnet06, White17, Yamazaki21}. 
Classical quantum chemistry calculations are performed with the PySCF package~\cite{pyscf} (the same hereinafter unless otherwise stated).
In our \TheMethod~simulation, we consider each hydrogen atom as a fragment. 
With both STO-3G and 6-31G basis set,
\TheMethod~is in excellent agreement with FCI result.
The coupled-cluster singles and doubles  (CCSD) method performs well near equilibrium bond distance; however,
in the dissociation limit (bond distance $>$ 1.7 Å) its calculation fails to converge~\cite{White17}.
For STO-3G basis set, conventional ESVQE with 20 qubits is performed for a limited number of bond distances 
due to the prohibitive computational cost.
It is found that in the dissociation limit \TheMethod~is more accurate than
conventional ESVQE despite the drastic reduction of the number of qubits.
This seemingly surprising outcome, along with a detailed analysis of the errors, are discussed in Appendix~\ref{appendix-error}.
To evaluate the error introduced by the limited basis set STO-3G,
we include results from the MRCI+Q+F12 method in the complete basis set (CBS) limit~\cite{White17},
which can be considered as the ground truth for the potential energy curve of \HChain,
By comparing Fig.~\ref{fig:H10}(a) and (b), we find that using larger basis set
brings the potential energy curve produced by \TheMethod~much closer
to the MRCI+Q+F12@CBS reference curve and the exact dissociation limit.

\begin{figure}
\centering
  \includegraphics[width=1.03\linewidth]{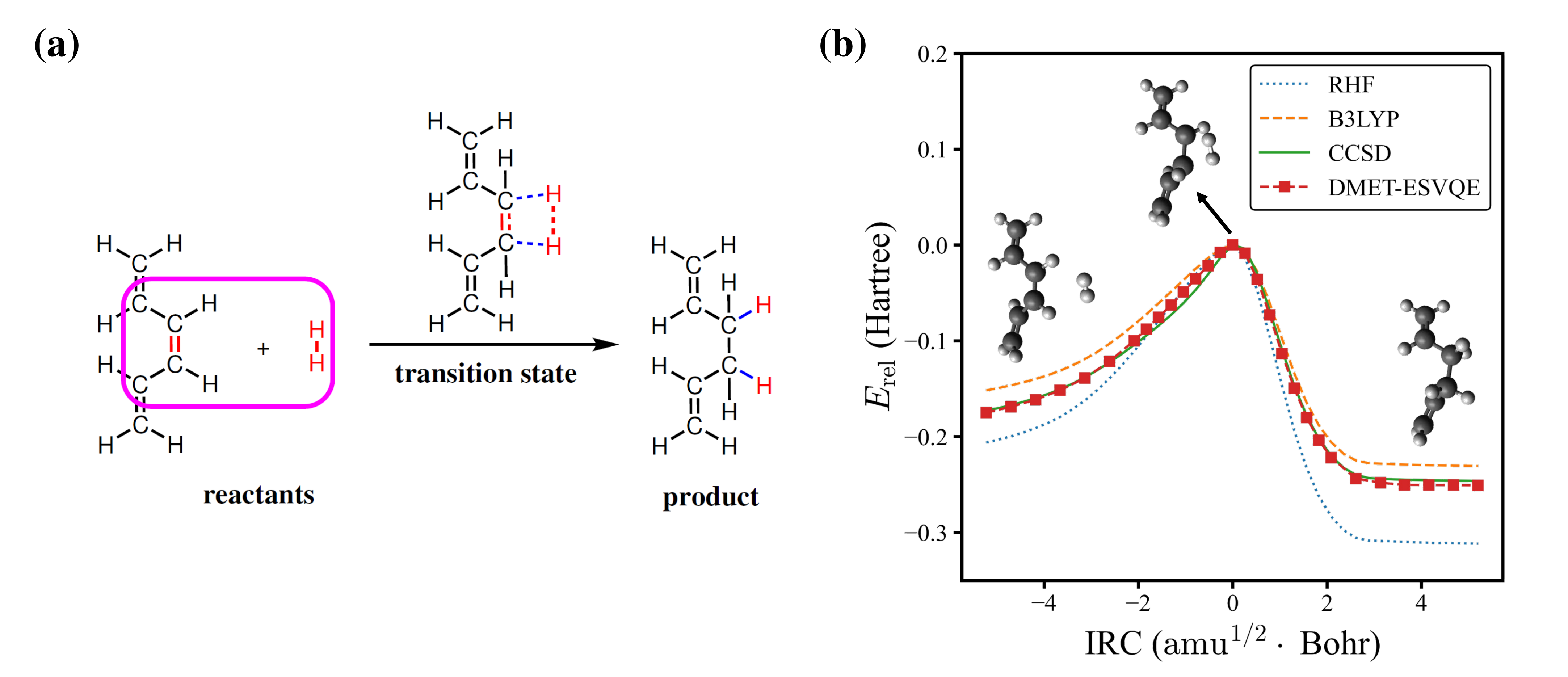}
  \caption{
  The potential energy curve for the hydrogenation reaction of \Alkene with \HM.
  (a) A schematic view for the hydrogenation reaction of \Alkene with \HM. 
  Each atom in the magenta box is considered as a single fragment.
  (b)
  Comparison of the energies obtained with RHF, B3LYP, CCSD and \TheMethod~along 
  the IRC
  of the reaction. The relative energy $E_{\rm{rel}}$ is $E - E_{\rm{TS}}$
  where $E_{\rm{TS}}$ is the transition state energy.
  Note that $E_{\rm{TS}}$ is different for different methods.
  }
  \label{fig:hydrogenation}
\end{figure}

Next, we study the energy profile for the addition reaction between
\Alkene and \HM \ in gas phase, 
which is a simplified model for the addition of hydrogen to conjugated hydrocarbons, an essential step for many organic synthesis routes~\cite{Robert14, Yao15Heavily, Pawlicki19}. A schematic diagram of the addition reaction is depicted in Fig.~\ref{fig:hydrogenation}(a).
A large fraction of the molecule is involved in conjugated $\pi$ bonds,
which poses a challenge for quantum embedding theories.
Besides, the transition state, 
which is defined as the first order saddle point in the potential energy surface, is known to be difficult for electronic structure methods.
In \TheMethod~simulation, each atom in the magenta box is considered as a single fragment
with the $1s$ orbitals for {carbon} atoms frozen.
The transition state and the intrinsic reaction coordinates~(IRC)~\cite{Fukui81} for the reaction are determined 
with hybrid DFT functional B3LYP under STO-3G basis set using the Gaussian 09 package~\cite{G09}.
In Fig.~\ref{fig:hydrogenation}(b), we plot the relative energy $E_{\rm{rel}} = E - E_{\rm{TS}}$
along with the IRC where $E_{\rm{TS}}$ is the transition state energy. The absolute value of $E_{\rm{TS}}$ can be found in the Appendix~\ref{appendix-ae}.
In agreement with common quantum chemistry perception,
it is observed that 
{restricted Hartree-Fock (RHF)} overestimates the reaction barrier, while B3LYP (DFT) underestimates the reaction barrier.
On the other hand, the energy profile generated by \TheMethod~is in remarkable agreement with the highly accurate and time-consuming CCSD method.
We note that  using basis set larger than STO-3G is essential for a more realistic description of the reaction.

\begin{figure}
\centering
  \includegraphics[width=\linewidth]{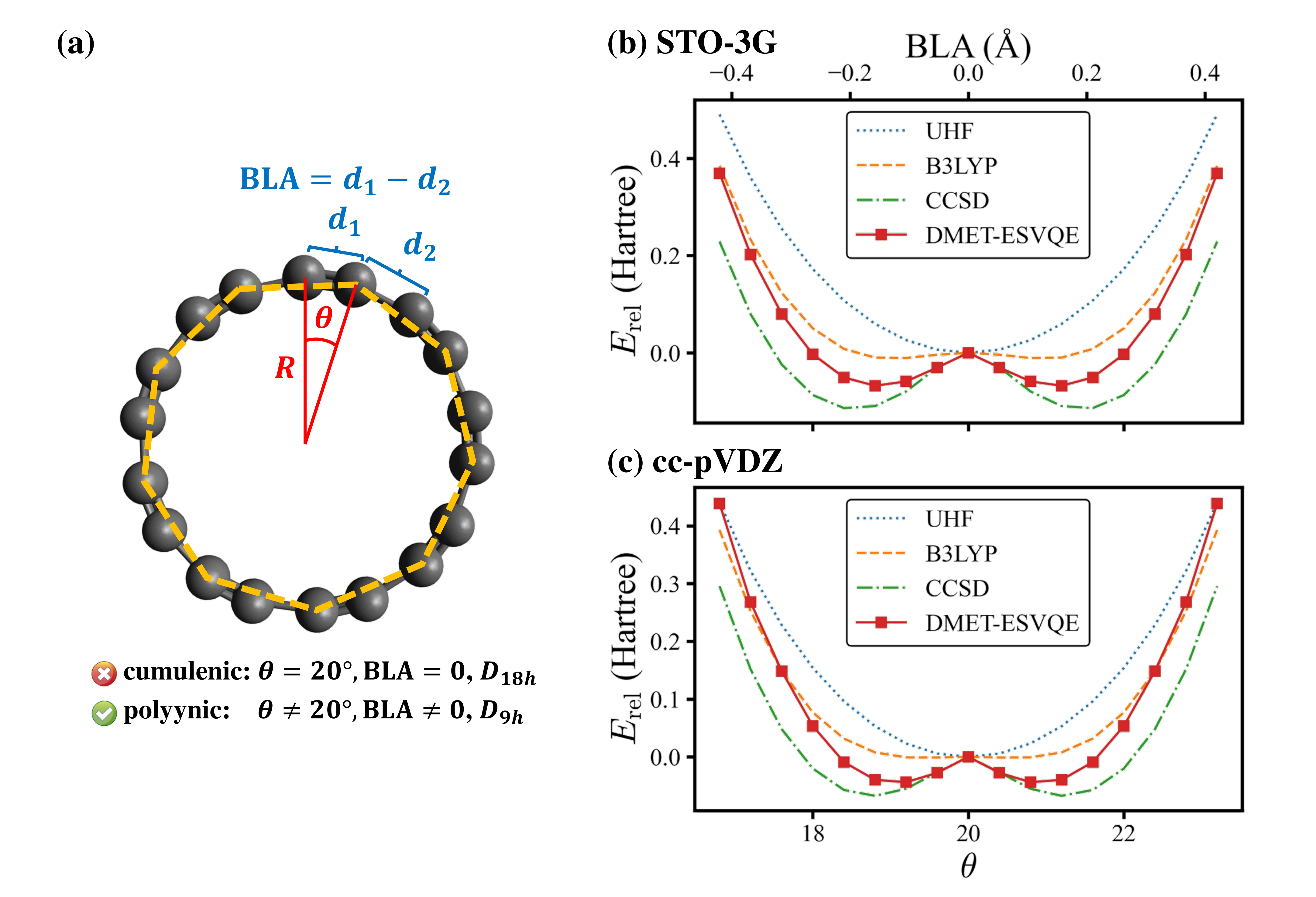}
  \caption{(a) A schematic diagram of the \CRing molecule. 
  $\theta$ is the angle between the two interleaving $\rm{C}_9$ nonagons,
  one of which is indicated with orange dashed lines.
  $R$ is the radius of the regular nonagons.
  $d_1$ and $d_2$ are the two sets of C-C bond lengths, respectively.
  (b)(c) Comparison of the energies obtained with UHF, B3LYP, CCSD and \TheMethod~for the
  potential energy curve of the \CRing molecule within (b) STO-3G and (c) cc-pVDZ basis set.
  The relative energy $E_{\rm{rel}}$ is defined as $E - E_{\rm{cumu}}$
  where $E_{\rm{cumu}}$ is the energy for the $\theta = 20 \degree$ cumulenic structure.
  The \TheMethod~results suggest that the bond-length alternating structure is favoured,
  which agrees with experimental observation.
  }
  \label{fig:C18}
\end{figure}

The last system studied is the \CRing molecule, a novel carbon allotrope with many potential applications such as molecular devices
due to its exotic electronic structure~\cite{Shen20, Mazziotti20, Boldyrev20, Qinxue20}. Before its experimental identification~\cite{Anderson19}, 
the equilibrium geometry of the molecule is under heated debate: DFT and perturbation theory (MP2) often conclude $D_{18h}$ cumulenic structure, yet high-level CCSD calculations indicate that bond-length and bond-angle alternated polyynic structure is more energetically favoured ~\cite{Ohno08}. 
In 2019, the polyynic structure is confirmed unambiguously 
via experimental synthesis of the molecule~\cite{Anderson19}.
In this work we investigate a series of geometries of \CRing molecule, shown in Fig.~\ref{fig:C18}(a), to determine the molecule's equilibrium geometry.
These geometries are generated by relatively rotating two interleaving $\rm{C}_9$ regular nonagons by an angle of $\theta \in [0, 40 \degree)$,
with all carbon atoms located on the same plane.
We define $d_1$ and $d_2$ as the lengths of the two sets of C-C bonds in the molecule and the bond length alternation~(BLA) as $d_1 - d_2$.
The $\theta = 20 \degree$ geometry is known as the cumulenic structure, while for other cases the geometries with $D_{9h}$ symmetry are called polyynic structure. 
The radius $R$ of the regular nonagon is determined
to be 3.824 Å via geometry optimization at CCSD/STO-3G level using
Gaussian 09 package~\cite{G09}.

In Fig.~\ref{fig:C18}(b), we present the potential energy curve
in the physically intriguing region $\theta \in [16.8\degree, 23.2\degree]$
within STO-3G basis set.
The relative energy $E_{\rm{rel}}$ is $E_{\rm{rel}} = E - E_{\rm{cumu}}$,
where $E_{\rm{cumu}}$ is the energy for the $\theta = 20 \degree$ cumulenic structure.
The absolute value of $E_{\rm{cumu}}$ can be found in the Appendix~\ref{appendix-ae}.
For this pathological system,
RHF is known to suffer from convergence problem~\cite{Qinxue20} and thus the {unrestricted Hartree-Fock (UHF)} results are shown. However, the UHF energy curve is qualitatively incorrect in that it anticipates the cumulenic structure to be more stable.
The representative DFT method B3LYP predicts rather flat potential energy curve around $\theta = 20\degree$ and the polyynic structure is slightly favoured by 11 mH compared to that of cumulenic structure. 
Because it is well documented that full degree of freedom optimization at the B3LYP level
yields cumulenic structure~\cite{Martin95,Ohno08, Qinxue20},
we believe the slight advantage of the polyynic structure shown in Fig.~\ref{fig:C18}(b) is an artifact of the fixed $R$.
In \TheMethod~simulation, we treat each carbon atom as a fragment with $1s$ orbital frozen.
Unlike UHF and B3LYP, \TheMethod~correctly reproduces the polyynic structure.
We note that solving the ground state of the full molecule with conventional VQE requires 144 qubits under frozen core approximation,
while for \TheMethod~16 qubits are sufficient for a correlated treatment of the whole molecule.
Fig.~\ref{fig:C18}(c) shows the results 
with Dunning's correlation-consistent basis set cc-pVDZ~\cite{Dunning89}.
In \TheMethod~simulation, the $2s$ and $2p$ basis orbitals for each carbon atom are considered as a single fragment and thus the effect of high angular momentum orbitals are treated at the mean-field level.
The general trends reflected by Fig.~\ref{fig:C18}(b) and Fig.~\ref{fig:C18}(c)
are consistent and
only CCSD and \TheMethod~are able to produce the correct equilibrium geometry.

\section{Conclusion}
In this work, we propose to integrate ESVQE with DMET for the study of realistic chemical problems. 
For benchmarking purpose, the typical model system \HChain \ is first tested with STO-3G basis set, and we find that \TheMethod~reaches near FCI accuracy.
DMET also enables ESVQE simulation of \HChain \ with 6-31G basis set, producing potential energy curve much closer to the reference result in the complete basis set limit.
The study of the hydrogenation reaction between \Alkene and \HM~shows that the accuracy of \TheMethod~is comparable to CCSD while the number of qubits required for VQE is reduced from 68 qubits to 16 qubits.
The last case studied in this work is the equilibrium geometry of the \CRing molecule
and it is found that \TheMethod~correctly predicts the experimentally observed polyynic structure with a significant reducetion of quantum resource, from 144 qubits to 16 qubits.
Our results suggest that the DMET embedding scheme
can effectively extend the simulation scale of the state of the art NISQ quantum computers.

To further expand the capability of the quantum embedding simulation, the effort could be divided into two directions: the first is related to the embedding scheme and the other is related to the high-level quantum solvers. For the embedding scheme part, the efforts could be further divided into three sub-directions. Firstly, one may need to develop more effective partition scheme to capture the correlation between the fragment and bath. Secondly, one may apply the self consistent fitting feature with a correlation potential to the DMET iteration and try other cost functions, respectively~\cite{Wouters16}. Particularly, one may consider speed up the convergence through
projected DMET~\cite{wu2019projected} and enhance the robustness and efficiency of DMET via semidefinite programming and local correlation potential fitting~\cite{wu2020enhancing}. Finally, one may consider the bootstrap embedding scheme, which has been tested on larger molecule system to achieve better accuracy and faster convergence~\cite{welborn2016bootstrap,ye2019bootstrap,ye2020bootstrap}.

For the high-level quantum solvers part, there are at least several directions that one can pursue. For the ESVQE part, one may reduce the energy threshold $\varepsilon$ to increase the operator pool size selected for the VQE iteration, thus further increasing the accuracy. Apart from the energy sorting scheme, it is worth trying other schemes such as $k$-UpCCGSD~\cite{lee2018generalized} to prepare the trial states in the high-level quantum solver. 
Note that if one jumps out of the current quantum-classical VQE framework, one may resort to the quantum imaginary time evolution solver, which could avoid high-dimensional parameter optimization~\cite{motta2020determining}. When implemented the high-level quantum solver on real quantum systems, one may explore quantum error mitigation methods to improve the accuracy of the measurement results~\cite{temme2017error,endo2018practical,strikis2020learning,sun2021mitigating,bravyi2021mitigating,kim2021scalable}. For more efficient simulation of larger molecule systems, advanced measurement schemes can be used to reduce the measurement cost, such as (derandomized) classical shadows or Pauli grouping methods to reduce the measurement overhead~\cite{huang2021efficient,kandala2017hardware,wu2021overlapped,o2016scalable,zhang2021experimental}.

The synergic development of quantum embedding theory, high-level quantum solver and quantum devices provides a great chance of solving strongly correlated chemical systems in future.
For example, one of the holy grails for quantum chemistry is the electronic structure of the iron-sulfur clusters of nitrogenase~\cite{reiher2017elucidating, zhendong19}, 
which contains eight transition metal atoms and exhibits strong correlation.
Within polarized triple-zeta basis set, each metal atom requires about 50 basis functions to describe.
If, in the future, 200 qubits with sufficiently long coherence time and high gate fidelity are available,
the clusters can be divided into fragments consisting of individual transition metal atom,
such that the fragment+bath problem  of the embedded transition metal atoms
can be solved accurately using efficient VQE algorithms.
The successful implementation of the proposed protocol may elucidate
the complicated interaction of the transition metal atoms
and push the boundary of theoretical chemistry.

\section*{Acknowledgement}
The authors gratefully thank
Jiajun Ren, Chong Sun, Hongzhou Ye, Hung Q. Pham, He Ma and
Xuelan Wen, Nan Sheng, Zhihao Cui, Zhen Huang,  and Ji Chen for helpful discussions and Hang Li for support and guidance.

\emph{Note added.---}
During the preparation of our manuscript, the work~\cite{mineh2021solving} appears recently, which uses DMET combined VQE to solve the Hubbard model.

\bibliographystyle{SciAdv}
\bibliography{reference.bib}

\widetext

\appendix

\section{VQE with Energy Sorting Unitary Coupled Cluster Ansatz}
\label{sec:VQE}
In this work, we use the energy sorting strategy to construct a more compact quantum circuit for the ground state searching.  
The workflow of ESVQE is summarized below:
\begin{enumerate}
    \item Generate the reference state, i.e. the Hartree-Fock state, and construct the operator pool $\mathcal{O}$ to build the wave function ansatz. For UCCSD ansatz, the operator pool $\mathcal{O}$ consists of all the possible single- and double-excitation operators $\hat{T}_{pr}$ and $\hat{T}_{pqrs}$ defined in Eq.~\ref{eq-excitation-operators}.
    \item VQE optimization iteration is carried out for each operator $\hat{T}_{i} \in \mathcal{O}$ for $E_i$. 
    The importance of the operator is evaluated by the energy difference with the reference state $\Delta E_{i} = E_{i}-E_{\textrm{ref}}$.
  $\Delta E_{i} = E_{i}-E_{\textrm{ref}}$ with $E_i = \min_{\theta_i} \braket{\Psi_{\textrm{ref}}|e^{-\theta_i (\hat T_i - \hat T_i^\dagger)}\hat{H}e^{\theta_i (\hat T_i - \hat T_i^\dagger)}|\Psi_{\textrm{ref}}} $ and $E_{\textrm{ref}} =   \braket{\Psi_{\textrm{ref}}| \hat{H} |\Psi_{\textrm{ref}}} $
  Then, sorted list $\mathcal{E}=\{(\Delta E_{i}, \hat{T}_{i})\}_{\textrm{sorted}}$ is formed. 
    
    \item  The operators with contributions above a threshold $|\Delta E_{i}| > \varepsilon$ are picked out and used to perform the VQE optimization. 
    In this work we set $\varepsilon = 1\times 10^{-5}$.
    
    \item  Extra fine-tuning can be performed by iteratively adding more operators to the ansatz until the energy difference $E^{(k-1)}-E^{(k)}$ between the ($k-1$)th and the $k$th iteration ($k \ge 1$) is smaller than a certain convergence criterion. In this work we skip this step for simplicity.
    
    \item Finally, output the circuit parameters corresponding to the optimized wave function $|\Psi^{\textrm{opt}}\rangle$ together with the energy $E^{\textrm{opt}}$ and exit.

\end{enumerate}

In the initialization process,  the reference energy  $E_{\textrm{ref}}$ in Step 2 on a Hartree-Fock state can be classically calculated efficiently. The energy $E_i$ is measured on a quantum computer, which is the additional measurement cost compared to conventional VQE.
In this work, we use the first-order Trotter decomposition.

\section{Density Matrix Embedding Theory}
\label{appendix:dmet}
In this section, we will first review the basics of the quantum embedding methods, following the discussions in previous works~\cite{Wouters16,DMET2012,DMET2013}. We then discuss the practical implementation for molecular systems. 

Density matrix embedding is one of the representative methods in the quantum embedding theory, first proposed in 2012~\cite{DMET2012,DMET2013}, which converts the original quantum system into a system composed of a fragment, the corresponding bath, and the pure environment.
The basic idea is to compress the dimension of the system by Schmidt decomposition of the wave function.
Imagine a Hilbert space composed of two orthonormal subspaces called fragment A with dimension $L_A$ and environment B with dimension $L_B$ ($L_A < L_B$).
the dimension of any wave function $|\Psi \rangle$ in this Hilbert space can be decomposed by Schmidt decomposition as

\begin{equation}
\begin{aligned}
|\Psi\rangle &=\sum_{i}^{L_{\mathrm{A}}} \sum_{j}^{L_{\mathrm{B}}} \Psi_{i j}|A_{i}\rangle|B_{j}\rangle =\sum_{i}^{L_{\mathrm{A}}} \sum_{j}^{L_{B}} \sum_{\alpha}^{L_{A}} U_{i \alpha} \lambda_{\alpha} V_{\alpha j}^{\dagger}|A_{i}\rangle|B_{j}\rangle =\sum_{\alpha}^{ L_{A}} \lambda_{\alpha}|\tilde{A}_{\alpha}\rangle|\tilde{B}_{\alpha}\rangle,
\end{aligned}
\end{equation}
where the states $|\tilde{B}_{\alpha}\rangle = \sum_{j}^{L_{B}} V_{\alpha j}^{\dagger} |B_{j}\rangle$ are deﬁned as the bath orbitals which are entangled with fragment orbitals $|\tilde{A}_{\alpha}\rangle = \sum_{i}^{L_{\mathrm{A}}} U_{i \alpha}|A_{i}\rangle$. 
If $|\Psi\rangle$ is the ground state of a Hamiltonian $H$, then it must also be the ground state of
\begin{equation}
    \label{eq-Hemb}
    \hat H_{\rm{emb}} = \hat P \hat H \hat P,
\end{equation}
which is the Hamiltonian for the embedded system composed of fragment plus its bath with the projector defined by
\begin{equation}
\hat P = \sum_{\alpha \beta}|\tilde{A}_{\alpha} \tilde{B}_{\beta}\rangle\langle\tilde{A}_{\alpha} \tilde{B}_{\beta}|.
\end{equation}
The spirit of DMET is that the solution of a small embedded system is the exact equivalent to the solution of the full system~\cite{Wouters16}, while the dimensions could be greatly reduced. However, the construction of $\hat P$ requires the exact ground state of the full system $\ket{\Psi}$ and thus the introduction of approximations is necessary. 
In practice, DMET algorithm is designed in a bootstrap manner. 
The mean-field approximation will be used for the full system to carry out the Schmidt decomposition and the embedded system will be solved by high-level method.
Some constraints will be introduced to regulate the high-level results for further improvement.

In the following, we first outline the procedure of \TheMethod~in Section~\ref{appendix:procedure}. Next, we discuss the construction of the embedding Hamiltonian in Section~\ref{appendix:DMET-construction} and the constraints in the practical implementation for realistic quantum chemistry problems in Section~\ref{appnedix:DMET-constraint}.

\subsection{Procedure of \TheMethod}
\label{appendix:procedure}
The entire process of calculating a chemical system by DMET-ESVQE is outlined as follows. 

\begin{enumerate}
    \item Partition the system into several fragments.
    
    \item Perform low-level, mean-field calculation on the entire system to obtain the ground state $|\Phi_0 \rangle$. 
    
    \item Select a fragment from the system, construct the corresponding bath from $|\Phi_0 \rangle$ by Schmidt decomposition. Construct the projector $\hat P$ and then obtain $\hat H_{\rm emb} = \hat P \hat H \hat P$ for the embedded system.
    
    \item Calculate the one-body~(${^1 D}$) and two-body reduced density matrix~(${^2 D}$) of the embedded system by ESVQE simulated on a classical computer or real quantum device in the future.
    
    \item Check if all the fragments have been traversed. If not, go back to step 3 and move to the next fragment.
    
    \item Check if the constraint has been satisfied. The different limitation has a different cost function ${\rm CF}$  (see details in Appendix~\ref{appendix:DMET-construction}). For the single-shot DMET, the global chemical potential $\mu_{\rm global}$ is introduced to conserve the electron number. If $\mathcal{L} \left(\mu_{\rm global}\right)$ is more than a settled threshold $\tau$, go back to step 3 with the optimized $\mu_{\rm global}$, and re-calculate all the fragments.
    \item Calculate expectations such as the total energy of the system democratically. 
\end{enumerate}
The pseudocode of the above DMET workflow is outlined in Algorithm~\ref{pseudo-code}.

\begin{algorithm}[H]
\label{pseudo-code}
\caption{Pseudocode for Density Matrix Embedding Theory}
\SetAlgoLined
Partition the full system with a given scheme  \;
$|\Phi_0 \rangle \leftarrow $ low-level method \;
$\mu_{\rm global} \leftarrow 0$ \;
$\tau \leftarrow 10^{-5}$ \;
\Do{$\lvert \rm{CF} (\mu_{\rm global}) \rvert$ $ > \tau$}{
    \For{\rm{fragment} $A$ $\in$ \textrm{system} }{
        Construct bath orbitals, $|B_{q}\rangle \leftarrow |\Phi_0 \rangle , |A_{p}\rangle$ \; 
        Build projection matrix, $\hat P \leftarrow |A_{p}\rangle, |B_{q}\rangle$ \;
        Obtain embedding Hamiltonian, $\hat H_{\rm emb} \leftarrow \hat H, \hat P,\mu_{\rm global}$ \;
        Get  ${^{1}D_A}$ and ${^{2}D_A}$ by ESVQE, ${^{1}D_A}, {^{2}D_A} \leftarrow \hat H_{\rm emb}, |A_{p}\rangle, |B_{q}\rangle$ \;
    }
    $ \text{CF}(\mu_{\rm global})$ and $\mu_{\rm global} \leftarrow N_{\rm occ}, \sum_{A} {^{1}D_A}$ \; 
}
Calculate observable expectation of interest $\leftarrow \sum_{A} {^{1}D_A}, \sum_{A} {^{2}D_A}$
\end{algorithm}

\subsection{Construction of embedded system in interaction formulation}
\label{appendix:DMET-construction}
In this subsection, we discuss the strategy proposed in Ref.~\cite{Wouters16}.
A straightforward approximation for the exact ground state is the low-level Hartree-Fock wave function. DMET uses this low-level wave function to construct the bath orbitals and solve the embedded system with a high-level solver.
The low-level wave function $|\Phi_0 \rangle$ obtained from the mean-field method could be written in second quantization as follows:
\begin{equation}
    |\Phi_{0}\rangle=\prod_{\mu \in N_\text {occ}} \hat{a}_{\mu}^{\dag}|\text{vac}\rangle,
\end{equation}
where $\{\hat{a}_{k}, \hat{a}_{k}^{\dag} | k \leq L \}$ is the set of the annihilation and creation operators on $L$ spin orbitals denoted by indices $k, l$. $N_{\rm occ}$ electrons are supposed to occupy the $N_{\rm occ}$ lowest spin orbitals denoted by index $\mu , \nu$. The mean-field state $|\Phi_0 \rangle$ is obtained under a selected basis set, of which the annihilation and creation operators are $\{ \hat{c}_{k}^{\dag}, \hat{c}_{k} | k\leq L]\}$. 
For convenience, all basis have been orthonormalized and localized.
In this work, we use the meta-löwdin method implemented in PySCF for this purpose~\cite{Garnet14, pyscf}, although other methods such as intrinsic atomic orbitals have been reported in the literature~\cite{Knizia13, DMET2013}. 
$\{ \hat{a}_{\mu}^{\dag} , \hat{a}_{\mu} | \mu \leq N_{\rm occ}\}$ and $\{ \hat{c}_{k}^{\dag}, \hat{c}_{k} | k \leq L\}$ are connected through a coefficient matrix $C$:
\begin{equation}
    \hat{a}_{\mu}^{\dagger}=\sum_{k = 1}^{L} \hat{c}_{k}^{\dagger} C_{k \mu},
\end{equation}
with the size of $L \times N_{\rm occ}$.
The one-body density matrix $^{1}D_{\rm mf}$ of the state is obtained as
\begin{equation}
{^{1}D_{\text{mf},kl}}=\langle\Phi_{0}|\hat{a}_{k}^{\dagger} \hat{a}_{l}| \Phi_{0}\rangle=\sum_{\mu}^{N_{\text {occ }}} C_{k \mu} C_{\mu l}^{\dagger}.
\end{equation}

For convenience, it is assumed that the orbitals of a selected fragment A are constructed by the first $L_A$ spin orbitals.
The $^{1}D_{\rm mf}$ could be written as 
\begin{equation}
    ^{1}D_{\rm mf}=\left[\begin{array}{ll}
    ^{1}D^{A}_{(L_A \times L_A)} & ^{1}D^{\rm inter}_{(L_A \times (L-L_A))} \\
    ^{1}D^{\dag \ \rm inter}_{((L-L_A) \times L_A)} & ^{1}D^{B}_{((L-L_A) \times (L-L_A))}
\end{array}\right],
\end{equation}

The environment submatrix ${^{1}D}^B_{\rm mf}$ constructed from ${^{1}D}_{\rm mf}$ can be decomposed as:
\begin{equation}
    {^{1}D}^B_{\rm mf}=\sum_{q}^{L - L_A} \lambda_{q}^{2}| B_{q}\rangle\langle B_{q}|,
\end{equation}
where $\lambda_q$ is the eigenvalue of the environment orbitals $| B_q \rangle$. The bath orbitals entangled with the fragment will contribute all the eigenvalues between 0 and 1 (or 2 if using spatial orbital), while occupied~(1 or 2) and unoccupied~(0) environment orbitals are separated from the embedded system, where the occupied environment orbital is named core orbital either. 
Due to MacDonald's theorem~\cite{PhysRev.43.830}, the bath orbitals will have the same dimension with the fragment. The projector $B$ for the environment, which has size $(L - L_A) \times (L - L_A)$, can be directly constructed from bath orbitals plus core orbitals. Thus we obtain the projector for the full system as:

\begin{equation}
    \hat P=\left[\begin{array}{ll}
I_{(L_A \times L_A)} & 0_{(L_A \times (L-L_A))} \\
0_{((L-L_A) \times L_A)} & B_{((L-L_A) \times (L-L_A))}.
\end{array}\right].
\end{equation}

Finally, we obtain the embedding Hamiltonian $H_{\rm emb}$ in the interacting bath formulation~\cite{Wouters16} directly using Eq.\ref{eq-Hemb}.
Based on $\hat H_{\rm emb}$,
the one-body reduced density matrix $^{1}D^{A}_{\rm high}$ for the embedded system: 
\begin{equation}
    ^{1}D^{A}_{\rm high}=\left[\begin{array}{ll}
^{1}D^{\text{frag},A}_{(L_A \times L_A)} & ^{1}D^{\text{inter},A}_{(L_A \times L_A)} \\
^{1}D^{\dag \ \text{inter},A}_{(L_A \times L_A)} & ^{1}D^{\text{bath},A}_{(L_A \times L_A)}
\end{array}\right],
\end{equation}
is obtained by a high-level quantum solver mentioned in Appendix~\ref{sec:VQE} and so is the two-body reduced density matrix $^{2}D^{A}_{\rm high}$.

\subsection{Constraint for high-level solution}
\label{appnedix:DMET-constraint}
After solving all fragments, some constraints could be introduced to regulate the high-level solutions self-consistently~\cite{Wouters16}.
The electrons would be re-distributed between the fragment and bath during the DMET iteration,
and as a result the number of electrons in the fragments may not sum up to the total number of electrons of the full system.

A global chemical potential $\mu_{\rm global}$ is introduced to fix this problem by modifying the $H_{\rm emb}$ as 
\begin{equation}
    \hat H_{\rm emb} \leftarrow \hat H_{\rm emb} - \mu_{\rm global} \sum^{L_A}_{r \in \Omega^A} \hat{a}^{\dag}_r \hat{a}_r.
\end{equation}

Here, we have defined $\Omega^A = \bigcup_{j} \Omega_{j}^A \setminus  \Omega^A_{\textrm{mf}}$,
where $\Omega^A_{\textrm{mf}}$ is a set of inactive orbitals treated at mean-field level and 
excluded from the DMET iteration.
We note $\Omega^A_{\textrm{mf}}$ could be an empty set.
The wavefunction in the fragment $A$ can be represented by
\begin{equation}
    \ket{\Psi^A} =  \ket{\Psi_{\Omega^A}} \otimes \ket{\Phi^A_{\textrm{mf}}},
\end{equation}
where $\ket{\Psi_{\Omega^A}}$ denotes the high-level wavefunction in the selected basis set $\Omega^A$, and $\ket{\Phi^A_{\textrm{mf}}}$ is the mean-field solution, a single product state spanning the basis of inactive orbital  $\Omega^A_{\textrm{mf}}$.
We can find that ${^{1}D}_{r r}^{\rm{mf}}$ is irrelevant to the self-consistency condition for DMET. Therefore, this method  is different from those simply adopting an active space high-level solver, in which the inactive orbitals will be involved in the optimization process. 
In our workflow, the Newton-Raphson method has been used to optimize $\mu_{ \rm global}$ by solving the equation  $\mathcal{L} \left(\mu_{\rm global}\right) = 0$.

As indicated in the main text, the single-shot DMET cost function is written as
\begin{equation}
\label{SM:eq-cf-mu}
    \mathcal{L} \left(\mu_{\rm global}\right)=\left(\sum_{A} \sum^{L_A}_{r \in \Omega^A} {^{1}D}_{r r}^{{\rm frag},A}\left(\mu_{\rm global}\right) + N_{\textrm{mf}} - N_{\rm occ}\right)^2 ,
\end{equation}
where
$N_{\textrm{mf}} = \sum_A \sum_{r \in \Omega^A_{\textrm{mf}} }{^{1}D}_{r r}^{\rm{mf}}$
is the number of electrons in the inactive orbitals
obtained at mean-field level and$ N_{\rm occ}$ is the total number of electrons.
The solution could be improved further by eliminating the discrepancy between the mean-field one-body density matrix and the fragment one-body density matrix by
adding a correlation potential $\hat{C}(u)$ to Hamiltonian $\hat H$:
\begin{equation}
    \hat H \leftarrow \hat H + \hat C(u),
\end{equation}
where $\hat C (u)$ takes the form
\begin{equation}
    \hat C(u) = \sum_{A} \sum_{r s \in \Omega^A} u_{rs} \hat{a}_r^{\dag} \hat{a}_s .
\end{equation}
The cost function can be written as
\begin{equation}
\mathcal{L}(u)=\sum_{A} \sum_{r s \in \Omega^A}\left({^1}D_{r s}^{\text{frag},A}-{^1}D_{r s}^{\mathrm{mf}}(u)\right)^{2} + \gamma \sum_{r s \in \bigcup_A \Omega^A_{\textrm{mf}}}\left({^1}D_{r s}^{\mathrm{mf}}(u)-{^1}D_{r s}^{\mathrm{mf}}(0)\right)^{2},
\end{equation}
where ${^1}D_{r s}^{\mathrm{mf}}(0)$ is the one-body reduced density matrix without the fitted correlation potential and $\gamma$ is a predefined weight constant. The last term ensures minimal effect of the correlation potential on the inactive orbitals $\Omega^A_{\textrm{mf}}$.

The approach which only keeps
the conservation of electron number is named as single-shot DMET~\cite{Wouters16},
while the one that introduces the correlation potential is named as correlation potential fitting DMET (or self-consistent DMET)~\cite{DMET2012,DMET2013,Wouters16}.

Single-shot DMET could also be modified case by case to improve the DMET performance or save the computational cost. In some cases, 
we are only interested in a small region in the whole system, such as the reaction center of a large organic molecule.
Then the interested region can be treated as the only fragment and the calculation of the embedded system is carried out without any constraint. This modification is named active space DMET or DMET(AS)~\cite{Wouters16}.
We note that from a chemical perspective, if FCI is used as the DMET solver, this exactly corresponds to a CASCI calculation and the role of DMET is to define an active space.

\subsection{Computation of expectation value}

The expectation will be calculated in a so-called democratic way, first proposed in Ref.~\cite{Wouters16}. It means if an operator has the indices from different fragments, the expectation of this operator will be the average of expectation for this operator in different fragments. For instance:
\begin{equation}
    \langle \hat{a}_{i}^{\dagger} \hat{a}_j + \hat{a}_{j}^{\dagger} \hat{a}_{i} \rangle= \langle\Psi_{A}| \hat{a}_i^{\dagger} \hat{a}_{j}| \Psi_{A}\rangle+\langle\Psi_{B}| \hat{a}_{j}^{\dagger} \hat{a}_{i}| \Psi_{B}\rangle,
\end{equation}
where the $i$ and $j$ belong to the fragment $A$ and $B$, respectively. So are the two-body terms and so on.
\section{Numerical results}

\subsection{Absolute energies}
\label{appendix-ae}
In Table.~\ref{tab:absolute-energies}, we list the absolute HF (either restricted or unrestricted), B3LYP, CCSD
and \TheMethod~energies for the specified geometries of \Alkene hydrogenation and the \CRing molecule.
All methods except HF are non-variational, so direct comparisons of the absolute energies are of limited significance.
 
\begin{table}
\caption{Absolute energies of the specified geometries
of the systems invested in the main text by HF, B3LYP, CCSD and \TheMethod. 
The meanings of the "Symbol"s can be found in the corresponding main text.} \label{tab:absolute-energies}
\addtolength{\tabcolsep}{+15pt}
\centering
\begin{tabularx}{\linewidth}{cccccc}
\hline \hline
System                & Symbol            & HF       & B3LYP     & CCSD     & DMET-ESVQE \\
\hline
\Alkene hydrogenation & $E_{\rm{TS}}$     & -229.853 & -231.361 & -230.371 & -230.275 \\
\CRing (STO-3G)       & $E_{\rm{cumu}}$   & -672.783 & -676.324 & -673.799 & -674.325 \\
\CRing (cc-pVDZ)      & $E_{\rm{cumu}}$   & -681.204 & -684.962 & -683.223 & -682.624               \\
\hline \hline              
\end{tabularx}
\addtolength{\tabcolsep}{-15pt}
\end{table}

\subsection{Analysis of the errors}
\label{appendix-error}
\begin{figure}
\centering
  \includegraphics[width=.62\linewidth]{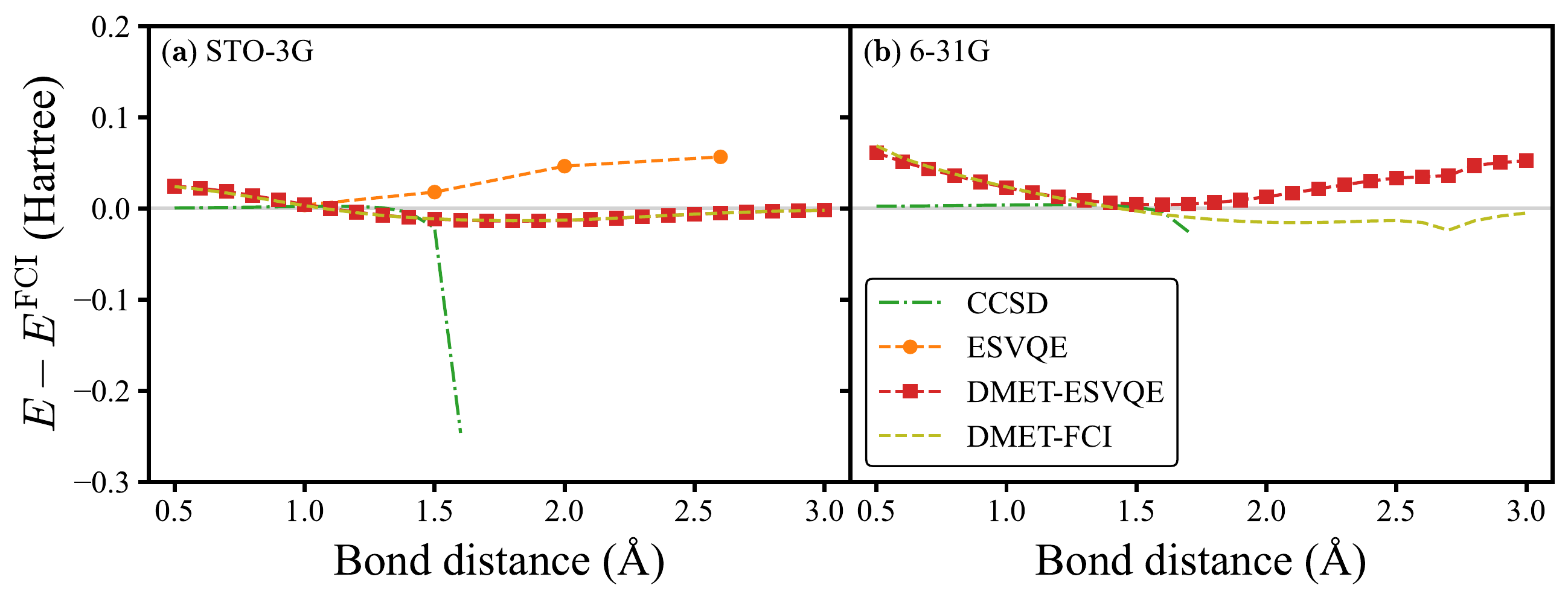}
  \caption{Relative error for the homogeneous stretching of a evenly-spaced hydrogen chain composed of 10 atoms in (a) STO-3G and (b) 6-31G basis set.
  }
  \label{fig:H10-error}
\end{figure}

Here we discuss the errors for \TheMethod~simulation in length 
to gain more insight into ESVQE and DMET.
In Fig.~\ref{fig:H10-error} we show the relative error for the results derived by 
CCSD, ESVQE and \TheMethod~in the \HChain~system with STO-3G and 6-31G basis set.
The reference ground truth is the results given by FCI in the respective basis set.
In both panel (a) and panel (b) the convergence failure for CCSD is clearly visible.
For STO-3G basis set, the results for conventional ESVQE
is worse then \TheMethod~proposed here. 
This outcome can be surprising at first glance
considering that the simulation of 
conventional ESVQE requires more qubits than \TheMethod.
Indeed, in the case where FCI is used as the high-level solver (DMET-FCI), 
DMET-FCI is apparently an approximation to the original FCI method
and the role of DMET is to reduce the computational cost.
The accuracy of DMET-FCI can be improved by using larger fragment size
until the fragment size is equal to half of the whole system.
However, the situation is not the same if approximate quantum chemistry solvers such as CCSD and ESVQE are considered.
In the specific case of \HChain~with each single H atom as a fragment, 
it is well established that DMET-FCI produces the exact result in the dissociation limit~\cite{Wouters16}.
The result is natural to understand in that DMET is effectively adding up energies of individual H atoms which is treated at the FCI level.
The key point of the \HChain~case is that DMET-CCSD or DMET-ESVQE is equivalent to DMET-FCI
because the fragment+bath problem is a two-electron two-orbital problem.
As a result, DMET-CCSD or DMET-ESVQE becomes more accurate than CCSD or ESVQE respectively at the dissociation limit.
Despite the argument here, we do not anticipate that the DMET framework is able to reduce computational cost and improve accuracy simultaneously for general chemical systems.
For 6-31G basis set, the error of \TheMethod~deviates from the error of DMET-FCI at the dissociation limit. The reason could be the energy sorting truncation of the ESVQE ansatz and the Trotter error~\cite{Mayhall20}.

\begin{figure}
\centering
  \includegraphics[width=.32\linewidth]{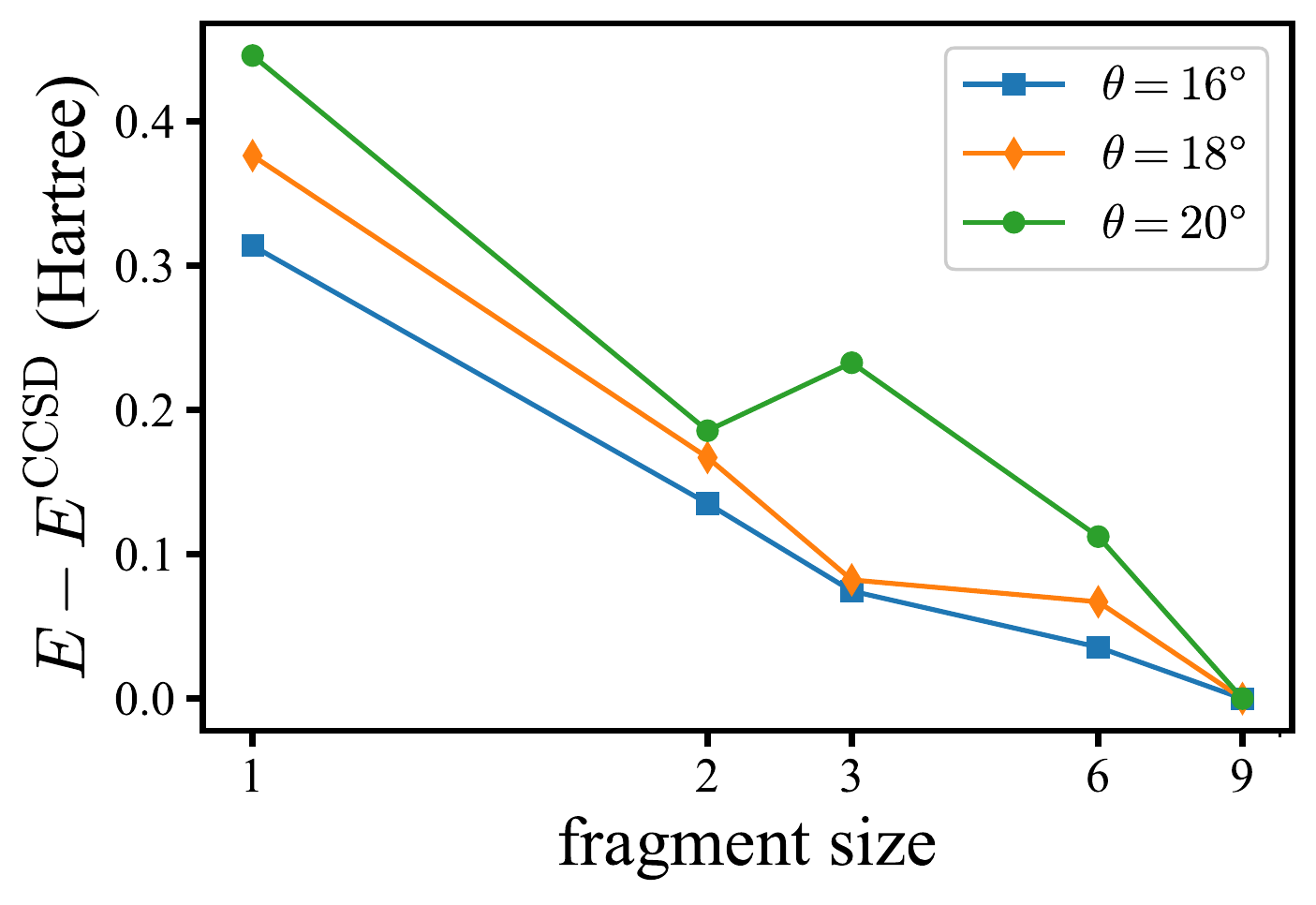}
  \caption{The convergence of DMET with respect to the fragment size using CCSD as the high level solver. The reference energy $E^{\textrm{CCSD}}$ is obtained by solving the whole \CRing molecule using CCSD. The fragment size is defined as the number of carbon atoms in each fragment.}
  \label{fig:C18-fragment}
\end{figure}

In Fig.~\ref{fig:C18} the potential energy curves by CCSD and \TheMethod~are not well aligned.
In Fig.~\ref{fig:C18-fragment} we explore its origin by checking
the convergence of fragment size with STO-3G basis set
using CCSD as the high-level solver, termed as DMET-CCSD.
Here we have replaced ESVQE with CCSD because their performance should be of the same level. 
The fragment size is defined as the number of carbon atoms contained in each fragment.
For the $\theta = 16\degree$ geometry and the $\theta = 18 \degree$ geometry
we find that by increasing the fragment size the energy obtained by DMET-CCSD
converges to the energy obtained by CCSD and 
$E - E^{\textrm{CCSD}}$ approaches to zero.
When the number of carbon atoms in the fragment is 9, DMET-CCSD effectively reduces to CCSD.
For the $\theta = 20\degree$ geometry, the difference between DMET-CCSD and CCSD abruptly increases when there are 3 carbon atoms in each fragment.
This is actually a known issue of DMET that using larger fragment size might deteriorate the outcome~\cite{Wouters16} and strategies to improve this shortcoming are under active development~\cite{welborn2016bootstrap}.

\end{document}